\documentclass[preprint]{aastex}
\usepackage{pdflscape}
\usepackage{afterpage}
\usepackage{capt-of}
\usepackage{rotating}
\usepackage{threeparttable}
\usepackage{amsmath}
\usepackage{amssymb}
\usepackage{url}
\usepackage{graphicx}
\newcommand{\be}{\begin{equation}}
\newcommand{\ee}{\end{equation}}
\newcommand{\bea}{\begin{eqnarray}}
\newcommand{\eea}{\end{eqnarray}}

\usepackage{color}              
\usepackage{epsfig}
\usepackage{amsmath}
\usepackage{amssymb}
\usepackage{amstext}

\begin{document}

\title{The peculiar cluster MACS J0417.5$-$1154 in the C and X$-$bands}
\author{Pritpal Sandhu\altaffilmark{1},  Siddharth Malu\altaffilmark{1}, Ramij Raja\altaffilmark{1} and Abhirup Datta\altaffilmark{1,2}}
\altaffiltext{1}{Centre of Astronomy, Indian Institute of Technology Indore, Simrol, Khandwa Road, Indore 453552, India}
\altaffiltext{2}{Center for Astrophysics and Space Astronomy, Department of Astrophysical and Planetary Science, University of Colorado, Boulder, C0 80309, USA}
\begin{abstract}
We present 5.5 and 9.0 GHz Australia Telescope Compact Array (ATCA) observations of the cluster MACSJ0417.5$-$1154, one of the most massive galaxy clusters and one of the brightest in X-ray in the Massive Cluster Survey (MACS).  We estimate diffuse emission at 5.5 and 9.0 GHz
from our ATCA observations, and compare the results with the 235 MHz and 610 MHz 
GMRT observations and 1575 MHz VLA observations. We also estimate the diffuse emission at low frequencies from existing GLEAM survey data (using the MWA telescope\footnote{http://www.mwatelescope.org}), and find that the steepening reported in earlier studies may have been an artefact of underestimates of diffuse emission at low frequencies. High-frequency radio observations of galaxy cluster mergers therefore provide an important complement to low-frequency observations, not only for a probing the `on' and `off' state of radio halos in these mergers, but also to constrain energetics of cluster mergers. We comment on the future directions that further studies of this cluster can take.
\end{abstract}


\section{Introduction}
\label{sec:intro}
Galaxy clusters, the most massive gravitationally bound objects in the
universe, grow through collisions and mergers. Galaxy cluster mergers
are characterized by cluster-wide magnetic fields, which, along with
accelerating charged particles in the intracluster medium (ICM), may give
rise to cluster-wide synchrotron emission (radio halos) as well as peripheral elongated shock-generated diffuse emission (radio relics). 

X-ray data traces the thermal component of the gas in the cluster, and radio emission traces the non-thermal components, which, in the case of cluster mergers, has its origins in the merger shock-induced particle acceleration, and magnetic field enhancement \citep{2014IJMPD..2330007B}. 

Diffuse emission in cluster mergers (i.e. both radio halos and radio
relics) has been well-studied in radio at a wide range of frequencies (see \citet{2012A&ARv..20...54F,2015FrASS...2....7M,2014IJMPD..2330007B} for reviews), and has yielded information about the dynamical states \& merger histories of, and magnetic fields in cluster mergers. In the
past decade, several cluster surveys, like the Massive Cluster Survey
(MACS) have characterized the dynamical states of several cluster
mergers -- data from these surveys, combined with increases in bandwidth and consequently sensitivities of
radio observatories (\citet{2011MNRAS.416..832W,2011ApJ...739L...1P}), have made the
detection of diffuse emission in several cluster mergers possible.

MACS J0417.5$-$1154 is a hot ($T\sim $11 keV), X--ray luminous ($L_{X}\sim $ 3.66$\times 10^{45}$ ergs s$^{-1}$) and one of the most massive ($M\sim 10^{15}M_{\mathrm{sun}}$, see \citet{2014MNRAS.439...48A}) clusters that was discovered by MACS \citep{2010MNRAS.407...83E}. The first radio observations were done with GMRT at 235 and 610 MHz by \citet{2011JApA...32..529D}, who were motivated by the peculiar nature of the cluster/merger, specifically the extremely flat spectrum of the diffuse emission in the cluster $-$ the flattest observed so far between 235 and 610 MHz \citep{2011JApA...32..529D}. In both \citet{2011JApA...32..529D} and \citet{2017MNRAS.464.2752P}, the cluster shows diffuse emission in its centre, that is co-located with the X-ray emitting hot gas, and has a south-east to north-west extension. The cluster centre has two unresolved point sources, detected at 235 and 610 MHz \citep{2011JApA...32..529D} and also at 1575 MHz, from Very Large Array (VLA) observations \citep{2017MNRAS.464.2752P}. Since the low-frequency end of diffuse emission in this cluster has been studied, we are interested primarily in pushing the high-frequency end. Our high-frequency radio observations constitute a complement to the work of \citet{2011JApA...32..529D} and \citet{2017MNRAS.464.2752P}. 

We present our observations in \S\ref{radioobs}, present the estimation of diffuse emission, and the spectrum thus obtained in \S\ref{diffem}, and discuss our results and future directions in \S\ref{discussion}.

\begin{figure}
\begin{center}
\includegraphics[width=\columnwidth,angle=0]{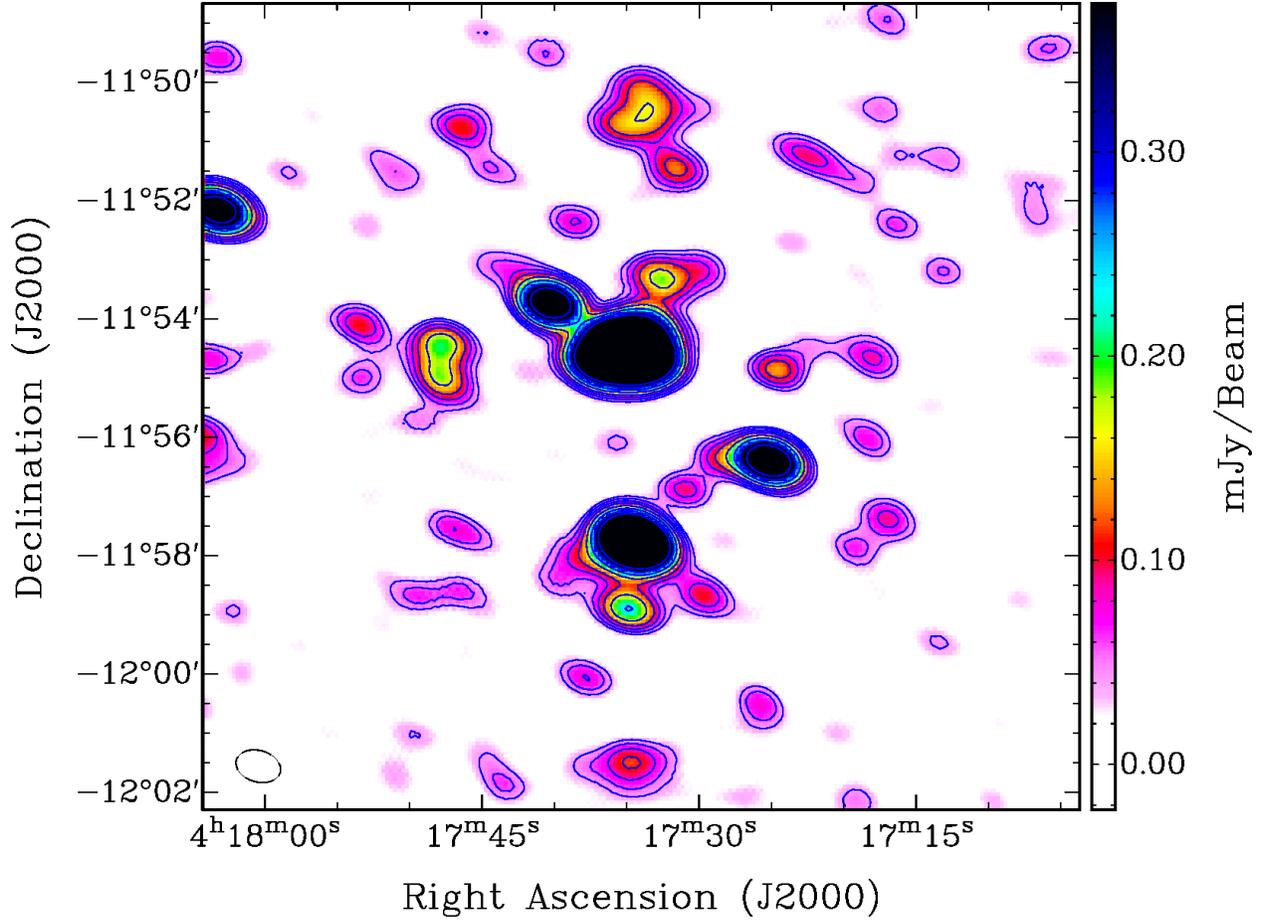}
\caption{A natural--weighted image of the entire region of the cluster
  MACS J0417$-$1154 at 5.5 GHz from ATCA. The beam, which is
  46.3$^{\prime\prime}\times$31.9$^{\prime\prime}$ with PA  74.02$^\circ$ is shown in the bottom
  left--hand corner. Noise rms is $\sigma$=8$\mu$Jy/beam. Contour levels start at 5$\sigma$, and increase by factors of $\sqrt{2}$. 
}\label{Fig1}
\end{center}
\end{figure}
 
\begin{figure}
\begin{center}
\includegraphics[width=\columnwidth,angle=0]{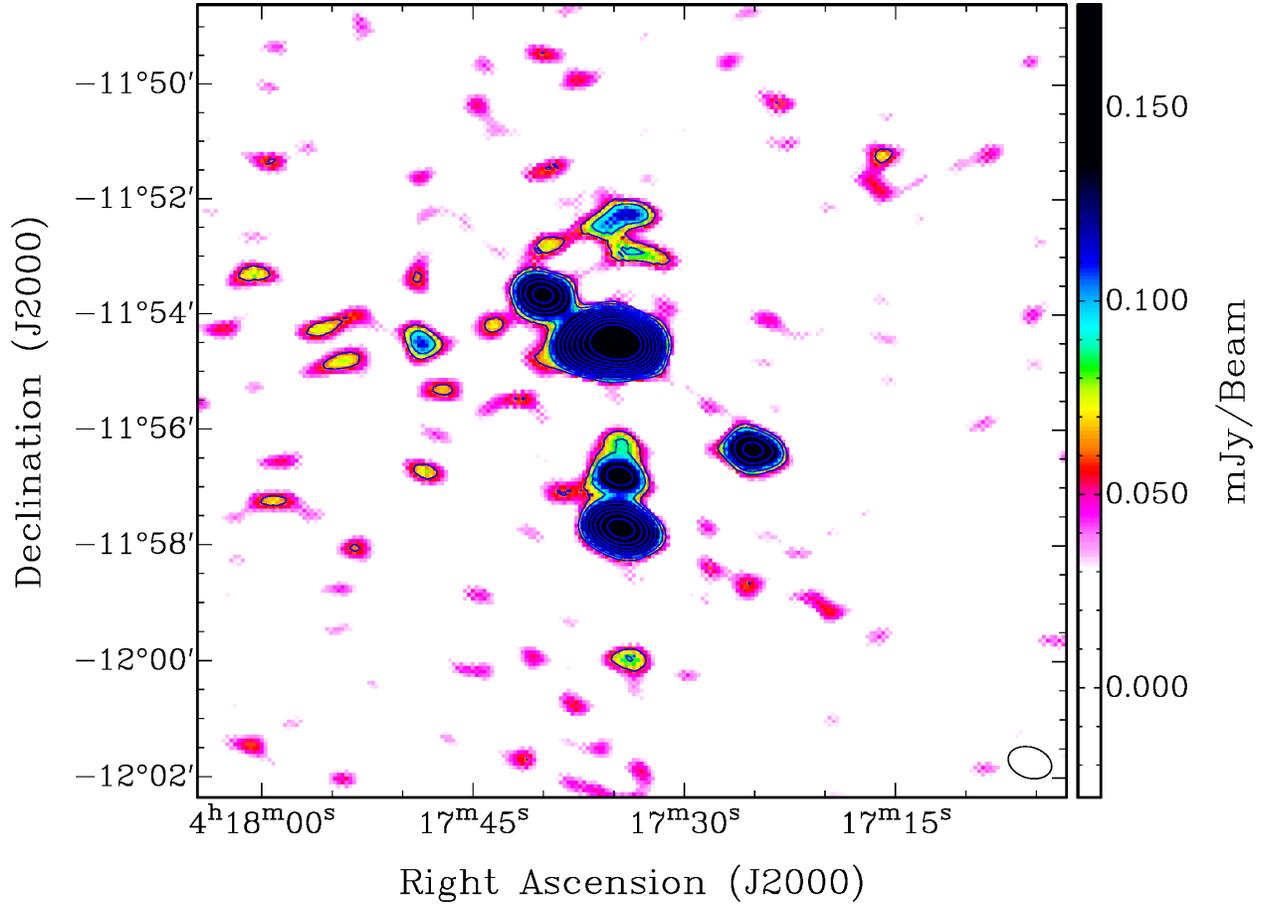}
\caption{A natural--weighted image of the central region of the cluster
  MACS J0417$-$1154 at 9 GHz from ATCA. The image has been convolved to the same beam as the 5.5 GHz image in Fig.(\ref{Fig1}), i.e. 46.3$^{\prime\prime}\times$31.9$^{\prime\prime}$ with PA  74.02$^\circ$, and is shown in the bottom
  left--hand corner. Noise rms is 12$\mu$Jy/beam. Contours start at
  5$\sigma$ (outermost contour) and increase by a factor of
  $\sqrt{2}$.}\label{Fig2}
\end{center}
\end{figure}

\begin{center}
\begin{table}
\caption{Summary of the ATCA observations}
\label{obs_journal}
\begin{tabular}{@{}llcr}
\hline
Array & Frequency & Observing    & Date \\
      & (GHz)     & time (hours) & \\
\hline
H168 & 5.5 & 8.0 & 2012 March 17 \\
H168 & 9.0 & 8.0  & 2012 March 17 \\
\hline
\end{tabular}
\end{table}
\end{center}

\section{Radio observations and Imaging}
\label{radioobs}
MACS J0417$-$1154 was observed for a total of 8 hours in the H168
array of the Australia Telescope Compact Array (ATCA), at 5.5 and 9.0 GHz. ATCA continuum mode with 1 MHz resolution \citep{2011MNRAS.416..832W}
with all four Stokes parameters and 2048
channels was used for these observations, in each of the two bands, centered at 5.5 and 9.0 GHz respectively. 1934$-$638 was used as the primary/amplitude
calibrator and 0403$-$132 as the secondary/phase calibrator. Data were
analyzed using Multichannel Image Reconstruction, Image Analysis and
Display (MIRIAD, developed by ATNF; see \citet{2011ascl.soft06007S,1995ASPC...77..433S} for details). Radio frequency interference induced bad data was excised. The
secondary/phase calibrator was observed once every 30 minutes to keep
track of variations in phase due to atmospheric effects. 

Our radio data reduction technique for ATCA data is described in \citet{2016Ap&SS.361..255M}; we summarize the self-calibration steps here. 
%
Self--calibration was done in three steps of phase
self--calibration and one of amplitude \& phase
self--calibration. A natural weighted image at 5.5 GHz is shown in
Fig.(\ref{Fig1}), and at 9 GHz in Fig. (\ref{Fig2}). 

A natural weighted image with about 4 times the size as Fig.(2) in \citet{2011JApA...32..529D} is shown in Fig.(\ref{Fig1}). The 9 GHz image in Fig. (\ref{Fig2}) has been convolved to the beam of the 5.5 GHz image in Fig. (\ref{Fig1}), i.e. 46.3$^{\prime\prime}\times$31.9$^{\prime\prime}$ with PA  74.02$^\circ$. 

\begin{center}
\begin{table*}
\caption{Unresolved and diffuse continuum radio sources detected in the central region of MACS J0417$-$1154}
\label{ptsrc}
\begin{tabular}{lcccccclrr}
\hline\hline
Source & \multicolumn{6}{c}{RA~~~~~~~~~~~~DEC} & Diffuse or
& {Int. Flux Density} & {Int. Flux Density} \\
Label & \multicolumn{6}{c}{(J2000)} & point source & $S_{\rm 5.5~GHz}$ & $S_{\rm 9~GHz}$ \\ \hline  & h & m & s & $^\circ$ & $\arcmin$ & $\arcsec$ &  &  (mJy) &  (mJy) \\ 
\hline
Source A & 04 & 17 & 34.7 & $-$11 & 54 & 30.6 & Point & 5.40 $\pm$ 0.70 & 2.40$\pm$0.30 \\
Source B & 04 & 17 & 36.7 & $-$11 & 54 & 38.7 & Point & 1.13 $\pm$ 0.03 & 0.71$\pm$0.07 \\
Central Region & 04 & 17 & 42.3 & $-$11 & 54 & 37.5 & Diffuse$+$Point & 7.76 $\pm$ 0.10 & 3.63 $\pm$ 0.10 \\ \hline\hline
\end{tabular}
\begin{tablenotes}[flushleft]
\item {\sc notes}-- Flux densities of sources in the central region of MACS J0417$-$1154. 
\end{tablenotes}
\end{table*}
\end{center}



\begin{figure}[!h]
\begin{center}
\includegraphics[width=\columnwidth,angle=0]{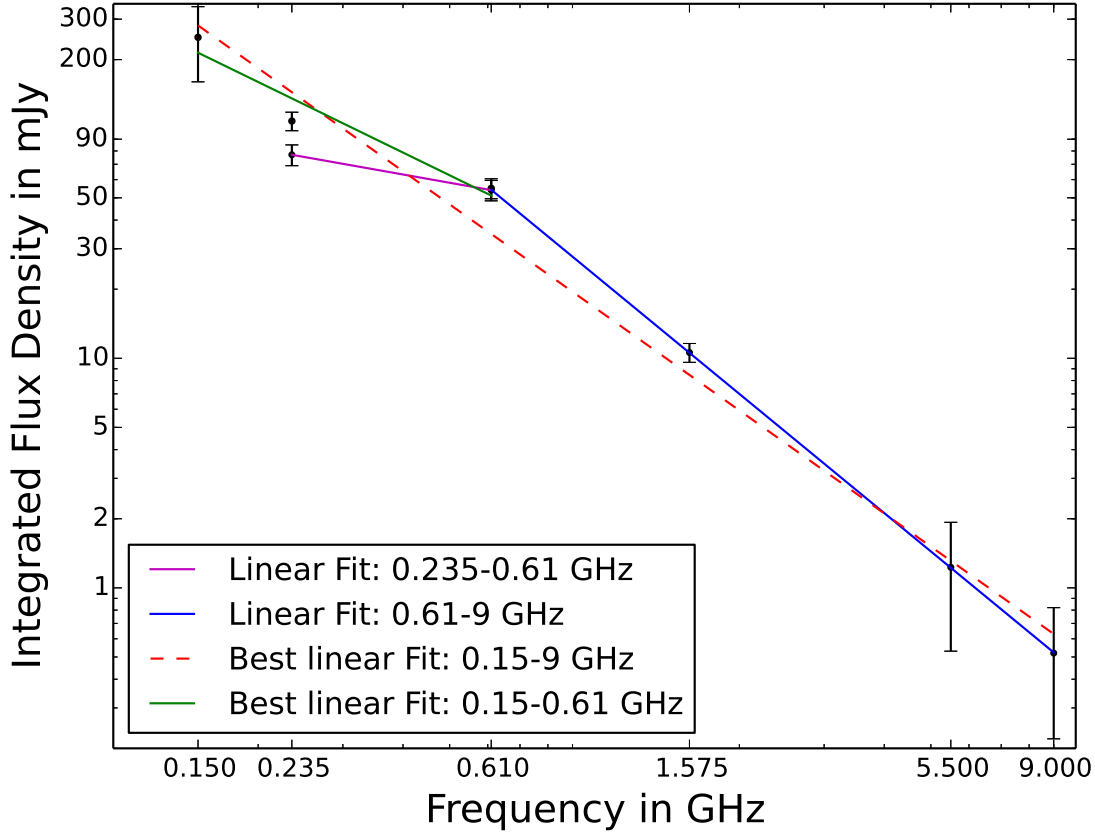}
\caption{The spectrum of diffuse emission in MACS J0417$-$1154, from 0.150 to 9 GHz. The best-fit spectral index is $\alpha= -$1.45, denoted by a dotted red line. The dotted red line is the single spectral index fit from 0.15 to 9 GHz, considering the upper limit on the diffuse emission at 0.235 GHz, as deduced from the GLEAM survey. Notice that a two-index fit, as shown through a green-line fit from 0.15 to 0.61 GHz, and a blue-line fit from 0.61 to 9 GHz fits the data better. Of the two data points at 0.235 GHz, the lower one is from earlier texts, i.e. \citet{2017MNRAS.464.2752P}, and the upper one is deduced from the GLEAM survey, as described in the text. The estimate at 0.15 GHz is deduced from the GLEAM and TGSS surveys, as described in the text.}\label{Fig7}
\end{center}
\end{figure}

\begin{figure}[!h]
\begin{center}
\includegraphics[width=\columnwidth,angle=0]{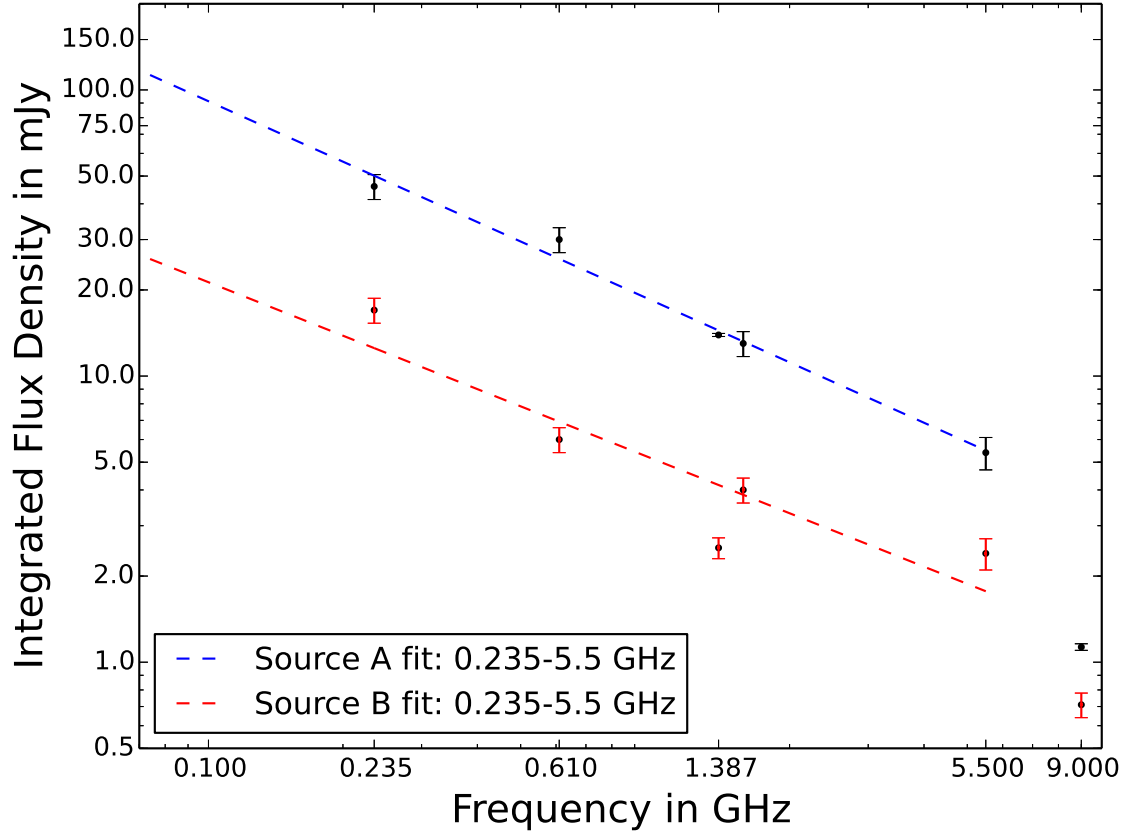}
\caption{The spectrum of the two point sources A and B. Black labels represent the source A and red ones represent B. Notice that the spectrum for both the sources steepens beyond 5.5 GHz. For this reason, the spectrum has been fit to a single spectral index up to 5.5 GHz. These single spectral index fits have been plotted down to 70 MHz.}\label{ptsrcspec}
\end{center}
\end{figure}

\begin{figure}[!h]
\begin{center}
\includegraphics[width=\columnwidth,angle=0]{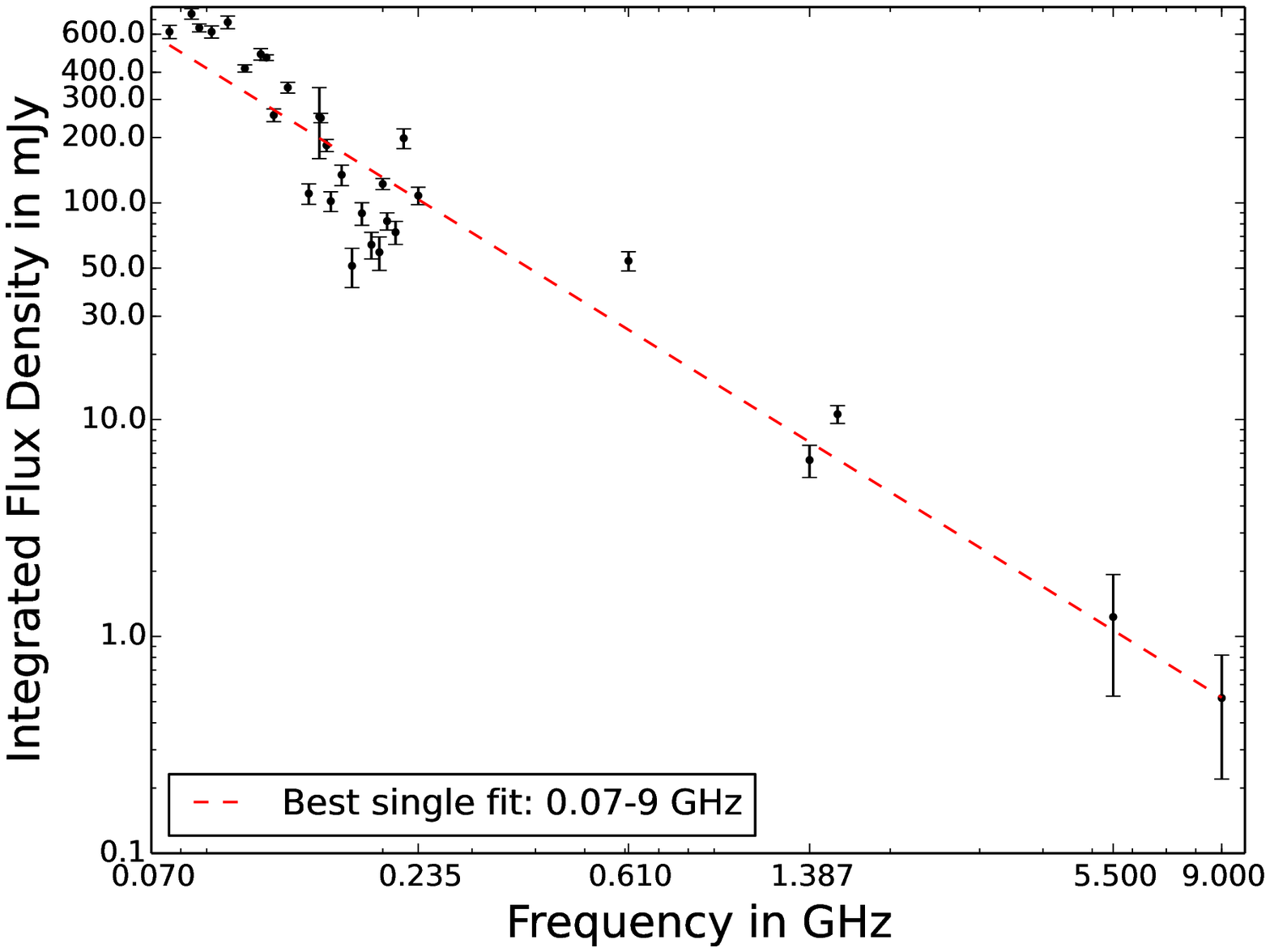}
\caption{The spectrum of diffuse emission at low frequencies estimated from GLEAM data.}\label{gleam1}
\end{center}
\end{figure}

\section{Estimation of Diffuse Emission at 5.5 and 9.0 GHz}
\label{diffem}
Figs. (\ref{Fig1}) and (\ref{Fig2}) are low-resolution, i.e. they do not have data from Antenna 6 of ATCA (at a distance of $\sim$ 3-6 km from the rest of the array, depending on the array configuration). Baselines involving Antenna 6 are able to provide a resolution of up to 5$^{\prime\prime}$ at 5.5 GHz and 2$^{\prime\prime}$ at 9 GHz. Being low-resolution images, it is not possible to estimate the amount of diffuse emission without data from Antenna 6. These low-resolution images are useful for estimating large-scale diffuse emission. 

In order to estimate the amount of diffuse emission in the cluster at 5.5 and 9 GHz, we adopted the following approach. 
\begin{enumerate}
\item First the total integrated flux of the central region, shown in Fig. (\ref{Fig1}) and Fig. (\ref{Fig2}) was estimated, at 5.5 and 9 GHz. To do this, the 9.0 GHz image was convolved to the beam size of the 5.5 GHz image, i.e. 46.3$^{\prime\prime}\times$31.9$^{\prime\prime}$ and PA 74.02$^\circ$. 
\item Then, images were made with a synthesized beam of 3$^{\prime\prime}$ at 5.5 and 2$^{\prime\prime}$ at 9 GHz (the best resolution available at the two frequencies respectively, {\bf using Antenna 6 data to resolve out the two sources}).
\item Integrated flux densities for the two sources, A and B (following labels from \citet{2017MNRAS.464.2752P}) were estimated by Gaussian fitting.
\item Finally, the fluxes of A and B were subtracted from the total integrated flux of the central regions obtained from the 46.3$^{\prime\prime}\times$31.9$^{\prime\prime}$ PA 74.02$^\circ$ images at 5.5 and 9.0 GHz. 
\end{enumerate}
Flux density of the diffuse emission in the central region of the cluster thus estimated is presented in Table \ref{ptsrc}, with details of the two point sources. 
\begin{center}
\begin{table*}
\caption{Diffuse emission in the central region}
\label{spectrum1}
\begin{tabular}{@{}lc}
\hline
Frequency & Diffuse Emission \\
 (MHz)    & (mJy)  \\
\hline
 235$^\ast$ & 77.0 $\pm$ 8.0 \\
 610$^\ast$ & 54.0 $\pm$  5.5\\
1575$^\dagger$ & 10.6 $\pm$ 1.0 \\
5500 & 1.23$\pm$ 0.7 \\
9000 & 0.52 $\pm$ 0.3 \\
\hline
\end{tabular}
\begin{tablenotes}[flushleft]
\item {\sc notes}-- $^{\ast\dagger}$These data points are quoted from \citet{2017MNRAS.464.2752P} and \citet{2011JApA...32..529D}. Data at 5500 and 9000 MHz above is estimated from the observations presented in this paper.
\end{tablenotes}
\end{table*}
\end{center}
\begin{figure}
\begin{center}
\includegraphics[width=\columnwidth,angle=0]{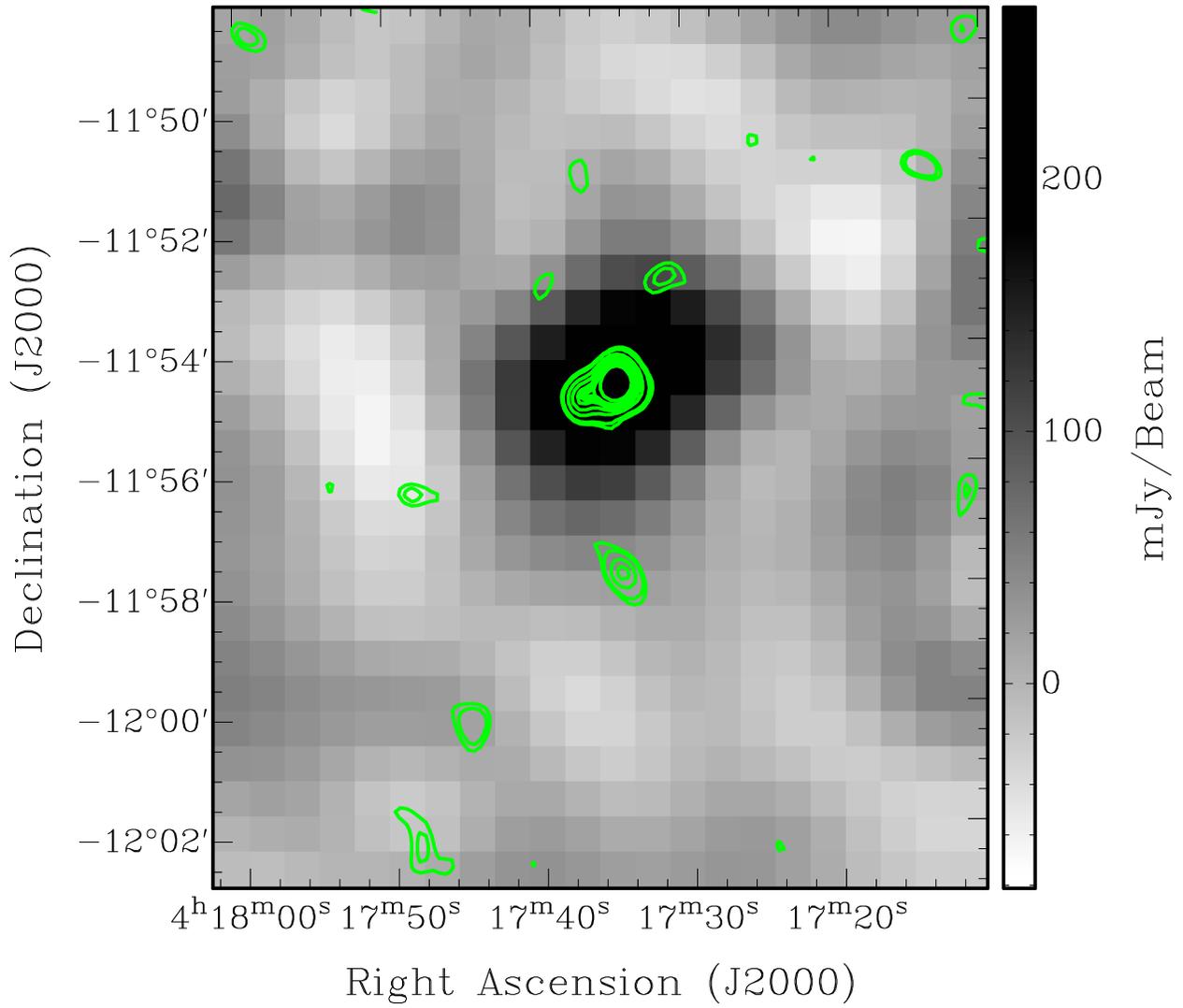}
\caption{TGSS contours overlaid on 150 MHz GLEAM image of MACS J0417.5$-$1154. Contour levels are (2.5, 3, 5, 7, 9, 10, 11, 15)$\times\sigma$, where $\sigma$ is 3 mJy/beam. }\label{Fig8}
\end{center}
\end{figure}

\subsection{Results}
\label{results}
Diffuse emission obtained after subtracting estimated fluxes of the sources `A' and `B' is presented in Table (\ref{spectrum1}) and as a spectrum, from 0.235 to 9.000 GHz, in Fig. (\ref{Fig7}). The calculation of spectral index from Table \ref{spectrum1} yields $\alpha_{9.00\mathrm{~GHz}}^{0.23\mathrm{~GHz}}$=1.45$^{+0.08}_{-0.06}$ -- this is the best-fit spectral index for the entire spectrum, from 235 MHz to 9 GHz. However, as shown in Fig. (\ref{Fig7}), two fits -- one from 0.235 to 0.61 GHz, and the other from 0.61 GHz to 9 GHz -- are a better fit to the data. Note from Fig. (\ref{Fig7}) that the spectrum is extremely flat at frequencies lower than 1 GHz, and then steepens very sharply, starting at 0.61 GHz. This sudden shift in the spectral index at 0.61 GHz that makes this a peculiar cluster. This sharp ``knee'', noticed by \citet{2017MNRAS.464.2752P}, at 610 MHz, yields a spectral index of $\alpha=-$1.72$^{+0.08}_{-0.10}$, which appears to be a good fit up to 9 GHz, from Fig. (\ref{Fig7}), and is consistent with the spectral index from 0.61 to 1.575 GHz from earlier low-frequency observations.

Alternatively, the radio image (Fig. \ref{Fig8}) from the GLEAM Survey \citep{2015PASA...32...25W,2017MNRAS.464.1146H}  
suggests that the reported 235 MHz diffuse emission flux may have been underestimated. We found from the GLEAM survey images at 223--231 MHz that the total intensity in the central region of the cluster is $\sim$ 171 mJy, and the total flux of the two sources, from \citet{2017MNRAS.464.2752P}, is $\sim$ 63 mJy. This would imply a total diffuse emission flux $\sim$ 108 mJy, which is significantly greater than the estimate of 77 mJy provided in \citet{2017MNRAS.464.2752P}. Unlike \citet{2017MNRAS.464.2752P}, we did find diffuse emission from TGSS \citep{2017A&A...598A..78I}. A total flux density of $\sim$ 154 mJy was found at 150 MHz in the central region, and a total flux density of $\sim$ 108 mJy was found for the two point sources A and B. We made estimates of point source fluxes of A and B at this frequency from the TGSS data and deduced the diffuse emission flux density, since the GLEAM survey provides images at 150 MHz as well. From the 150 MHz image, we deduced a total emission of $\sim$ 360 $\pm$ 90 mJy from the central region. This implies that the flux density of diffuse emission at 150 MHz is $\sim$ 250 mJy. 
These observations at 150--235 MHz from GLEAM may mean that the `knee' in the spectrum at 610 MHz is not as dramatic as it appears to be, and that the sharp cutoff at 610 MHz is at least in part due to the underestimation of diffuse emission at 235 MHz. 

As can be seen in Fig. \ref{Fig7}, the entire spectrum from 150 MHz to 9 GHz needs two-spectral-index fit, but the change in the spectral index is no longer sharp. Instead of the spectral index of $-0.37$ between 0.235 and 0.61 GHz, we now get a spectral index of $-1.04^{+0.35}_{-0.26}$ between 0.15 and 0.61 GHz, a change by a factor $>$2. The change to a spectral index of $-1.72^{+0.08}_{-0.10}$, for 0.61 to 9 GHz, is significantly less dramatic. 

Additionally, the spectrum of diffuse emission from lower frequencies has also been presented, as shown in Fig. (\ref{gleam1}). In order to obtain the diffuse emission at frequencies ranging from 76 MHz to 222 MHz, the following approach was adopted. First, a single-spectral index was fitted to the spectrum of sources A and B, which is shown in Fig. (\ref{ptsrcspec}). This spectrum was then extrapolated to the frequency range 76--222 MHz, for both sources A and B. The total calculated flux for A and B was then subtracted from the measured flux of the central region of the cluster at these frequencies, from GLEAM data. In fitting the spectrum of the sources A and B to a single spectral index, frequencies up to 5.5 GHz were considered. This is because of the steepening of the spectrum, as can be seen in Fig. (\ref{ptsrcspec}) at 5.5 GHz. Since we are interested in the low-frequency end of the spectrum, we do not include the 9 GHz data point while estimating the spectral indices, which are found to be $-0.70\pm0.07$ and $-0.62\pm0.07$ for the two sources A and B respectively. Fig. (\ref{ptsrcspec}) shows these single spectral index fits for the two sources A and B, extended to 70 MHz.

Using the above method, the spectrum of the diffuse emission thus estimated, as shown in Fig. (\ref{gleam1}), can be fit with a single spectral index $\alpha=-1.45\pm0.08$, between 76 MHz and 9 GHz. This is not significantly different from the value of the spectral index estimated between 235 MHz and 9 GHz, as above. This shows that it is possible to fit a single spectral index to the entire dataset, from 76 MHz to 9 GHz. As can be seen from Fig. (\ref{gleam1}), there are a few data points from the GLEAM data, for which diffuse emission may have been underestimated -- using any of these data points, instead of the entire GLEAM dataset, may well lead to an appearance of spectral steepening. 

Our estimates of diffuse emission in the cluster indicate that the spectrum at higher radio frequencies is consistent with the steep spectrum up to the L band, reported earlier \citet{2017MNRAS.464.2752P}; that is, there is no break in the spectrum from 1.575 to 9.000 GHz. Additionally, the estimates of diffuse emission we made from the GLEAM data further Considering this is one of the most massive and X-ray luminous clusters in MACS, it is curious that it contains little diffuse emission, indicating low activity in radio. 

\section{Discussion and Summary}
\label{discussion}
We have presented 5.5 and 9.0 GHz observations of MACS J0417.5$-$1154, a hot, X-ray luminous, massive
galaxy cluster with disturbed X--ray contours, using
the H168 array of the ATCA, and found that diffuse
emission is present at these two frequencies. Our data is able to constrain the diffuse emission present in the cluster, at 5.5 and 9.0 GHz.

This cluster -- which was earlier known to host a radio halo with one of the flattest spectra known -- shows the peculiarity of one of the sharpest `turn-off' point or `knee' in the spectrum, at 610 MHz. Such `knees' have been observed in other clusters \citep{2008Natur.455..944B}, though the sudden transition from such a flat spectrum to a steep one, is interesting. 

We point out that the images in \citet{2017MNRAS.464.2752P}, with a resolution of 20$\arcsec\times$20$\arcsec$, do not show any structure for the two sources `A' and `B'. It is, however, not clear what fraction, if any, of the diffuse emission in the central region is to be attributed to the two compact sources -- which may be radio galaxies -- and what fraction is part of the radio halo. Moreover, the spectral index in both radio halos and radio relics demonstrates large variations across the cluster. Unless the two sources are studied closely at multiple wavelengths, with high-resolution data, it may not be possible to make definite comments about the source of the diffuse emission in this cluster. Since we have subtracted the flux obtained at and in the vicinity of sources `A' and `B', our estimates of diffuse emission at 5.5 and 9.0 GHz are really lower limits. 

While recent literature on the subject \citep{2015ApJ...815..116F,0004-637X-809-2-186} discusses spectral curvature, and how it is affected by the morphology of non-thermal particles, this spectrum does not indicate any curvature, but just a two-spectral index fit, with the second spectral index yielding an excellent fit to the data up to 9 GHz. In this context, \citet{2016ApJ...823...13K,0004-637X-823-1-13,0004-637X-809-2-186} have considered a cluster merger shock that passes through a small region containing fossil non-thermal electrons. They show that in such a scenario, the spectrum would be significantly steeper than expected through radiative cooling. Kang and Ryu (2015) consider such a model with the small region of fossil non-thermal particles elongated perpendicular to the direction of the shock, as a way to reproduce the spectral curvature of the ``Sausage'' radio relic \citep{2014MNRAS.441L..41S,2016JKAS...49..145K}. Even though their model pertains to relic formation, it may be possible to use a model with a different morphology of the fossil non-thermal particles, in order to explain the hard break in the spectrum of MACS J0417$-$1154 at 610 MHz. 

\citet{2016ApJ...823...13K} point out the role of turbulent re-acceleration of cosmic ray (CR) electrons, which have been weakly accelerated by the cluster merger shock. They show that strong turbulence behind the shock can accelerate the CR electrons -- this may be a possible reason for the spectral index continuing without a break from 0.61 to 9 GHz. A break at 0.61 GHz in this model may indicate a less energetic cluster merger.

The sudden steepening of the spectral index may be indicative of spectral aging, but the steepened spectrum extending from 0.61 to 9.00 GHz (i.e. more than an order of magnitude in frequency) is, to the best of our knowledge, unprecedented. This sharp change suggests that the integrated flux of this radio halo should be measured at frequencies between 235 and 1575 MHz, to study the steepening of the spectrum, and above 10 GHz, to study spectral curvature. 

%
However, we have shown, through archival GLEAM and TGSS data, that the flux of diffuse emission may have been underestimated in earlier studies. This changes the behaviour of the spectrum considerably, from one that exhibits a sharp turnover at 0.61 GHz, to one that exhibits either a gentler steepening, or none at all, as can be seen from Fig. (\ref{gleam1}). This gentle steepening may not require a unique, exotic explanation, and is likely a signature of a relatively new/young galaxy cluster merger, which seems to be the case here. 

As mentioned earlier, diffuse emission may have been underestimated for a few of the low-frequency GLEAM data points, in the range 170 MHz to 200 MHz -- choosing data points in the range 76-235 MHz selectively may therefore lead to an appearance of steepening. In order to reliably remove the contribution of the two sources A and B, high-resolution data of the central cluster region is needed at these frequencies. This is where new low frequency surveys can play a critical role in confirming the spectral characteristics of the diffuse emission in this cluster. 

In conclusion, we have shown, through data at 5.5 and 9 GHz from ATCA, that the cluster merger MACS J0417$-$1154 has several peculiarities in the synchrotron emission from its centre, and that these peculiarities point to a need to study both effects of fossil non-thermal electrons lurking in one of the clusters before the merger, as well as the role of turbulent re-acceleration in the formation of radio halos. 

The aforesaid discussion assumes that the diffuse emission in the central region of this cluster is part of a radio halo, due to a cluster merger. However, the two sources in the central region may not be compact, and may have some diffuse emission associated with them, a detail that cannot be deduced from a simple Gaussian modeling of the sources, as in \citet{2017MNRAS.464.2752P}. {\bf Additionally, diffuse emission at 235 MHz is likely underestimated, from the GLEAM survey data. This implies that the break in the spectrum is likely not as sharp as it appears to be}. For these reasons, we stress the need for high-resolution multi-frequency images of the central region of this cluster, at frequencies in the range 0.2--1.57 GHz, in order to characterize the variation of the spectral index of diffuse emission. 

Thus, the new low frequency surveys from MWA, GMRT and LOFAR are critical for a well-calibrated study of diffuse emission in cluster mergers. 

\section*{Acknowledgements}
The Australia Telescope Compact Array is part of the Australia
Telescope which is funded by the Commonwealth of Australia for
operation as a National Facility managed by CSIRO. 
Analysis and observations were made possible by 
a generous grant for Astronomy by IIT Indore. GLEAM Postage Stamp Service: The GaLactic and Extragalactic MWA Survey Postage Stamp Service was used for this paper. We are grateful to the anonymous referee for their comments, which helped improve this paper. 

\bibliographystyle{spr-mp-nameyear-cnd}
\bibliography{bullet91}

\end{document}